\documentclass[journal]{IEEEtran}
\usepackage[pdftex]{graphicx}
\ifCLASSINFOpdf
\else
\fi

\usepackage{multirow}
\usepackage{amsmath}
\usepackage{amsfonts}
\usepackage{caption}
\usepackage{subcaption}
\usepackage{xcolor}

\usepackage{hyperref}
\usepackage{url}            
\begin{document}
%
\title{
Deep representation of EEG data from Spatio-Spectral Feature Images
}

\author{Nikesh~Bajaj$^{1,2}$,
        Jes\'us~Requena~Carri\'on$^{1}$, 
        and~Francesco~Bellotti$^{2}$%
        \\{\it {\small $^1$School of Electronics Engineering and Computer Science, Queen Mary University of London, UK}}
        \\{\it {\small $^2$Dipartimento di Ingegneria navale, elettrica, elettronica e delle telecomunicazioni, University of Genoa, Italy}}
}

\markboth{{\it \MakeLowercase{\href{https://phyaat.github.io}{\textit{https://phyaat.github.io}}}}}%
{N.Bajaj \MakeLowercase{\textit{et al.}}: IEEEtran}
%



\maketitle

\begin{abstract}
Unlike conventional data such as natural images, audio and speech, raw multi-channel Electroencephalogram (EEG) data are difficult to interpret. Modern deep neural networks have shown promising results in EEG studies, however finding robust invariant representations of EEG data across subjects remains a challenge, due to differences in brain folding structures. Thus, invariant representations of EEG data would be desirable to improve our understanding of the brain activity and to use them effectively during transfer learning. In this paper, we propose an approach to learn deep representations of EEG data by exploiting spatial relationships between recording electrodes and encoding them in a Spatio-Spectral Feature Images.  We use multi-channel EEG signals from the PhyAAt dataset for auditory tasks and train a Convolutional Neural Network (CNN) on 25 subjects individually. Afterwards, we generate the input patterns that activate deep neurons across all the subjects. The generated pattern can be seen as a map of the brain activity in different spatial regions. Our analysis reveals the existence of specific brain regions related to different tasks. Low-level features focusing on larger regions and high-level features focusing on a smaller and very specific cluster of regions are also identified. Interestingly, similar patterns are found across different subjects although the activities appear in different regions. Our analysis also reveals common brain regions across subjects, which can be used as generalized representations. Our proposed approach allows us to find more interpretable representations of EEG data, which can further be used for effective transfer learning. 

\end{abstract}


%
\IEEEpeerreviewmaketitle

%
%
%
%

\section{Introduction}
\label{S:Intro}
Brain-Computer Interface (BCI) systems are becoming increasingly popular. The ease of recording Electroencephlogram (EEG) signals has facilitated devising and launching new BCI systems for day-to-day applications, ranging from medical uses \cite{wolpaw2002brain} to gaming \cite{bellotti2013assessment}.
However, BCI systems that are trained on EEG signals from one subject might not perform well when applied to other subjects. This difficulty is generally ascribed to individual differences in the brain folding structure, which can result in EEG signals that follow different distributions.
Consequently, BCI systems might need to be re-trained on each future subject, which requires collecting and processing new EEG data and is a time-consuming activity. 

Several frameworks that use the principles of transfer learning have been proposed to calibrate pre-trained EEG-based systems, 
including filter banks \cite{tu2012subject}, adaptive feature extraction \cite{sun2014review}, transfer component analysis \cite{zhang2015transfer}, common spatial pattern \cite{kang2009composite, devlaminck2011multisubject, samek2013transferring}, regularised covariance matrix \cite{lotte2010learning, kang2009composite},  canonical correlation analysis \cite{yuan2015enhancing} and Convolutional Neural Networks (CNN) \cite{volker2018deep}. The time-consuming nature of the calibration process remains a major obstacle, 
and hence, some strategies have been developed to reduce the calibration time on new subjects \cite{dalhoumi2014knowledge, lotte2010learning}. 

Transfer learning and re-calibration in EEG studies can be improved by identifying common EEG patterns across subjects, i.e. invariant features \cite{wan2021review}.
However, unlike conventional data such as images of physical objects, speech signals or text, raw EEG data cannot be easily interpreted. A limited number of studies have focused on learning effective and robust feature representations of EEG data, which has the potential to be used not only for transfer learning, but also for explaining the decision process of a model and for understanding the brain mechanisms. In \cite{stober2015deep, nurse2015generalizable}, the weights learned in deep neural networks trained on EEG signals were investigated. Spatio-temporal representations of EEG signals were used as input for 2D CNN in \cite{zhang2017eeg}, and \cite{nurse2015generalizable} used multi-channel EEG signals as input images to a DNN to discover new features \cite{nurse2016decoding}. Spectral domain representations of EEG signals have been widely used in neuroscience and EEG studies \cite{sanei2013eeg} and therefore using spectral information in combination with spatial information resulting from multiple recording sites, can be an effective way of obtaining deep representations of EEG signals, a few studies have used spatio spectral filters for EEG studies \cite{lemm2005spatio}. 

There are a very few studies exploring the deep representation of EEG data. In \cite{bashivan2015learning} spectral and spatial information of EEG signals were combined in a 3-channel RGB image and used for training a DNN, from which deep representations were obtained. A few studies use transfer learning based on deep representation \cite{zhao2020deep} \cite{tan2018deep}, however the objective in such studies is the domain adaption. 

In this paper, we propose an approach for generating deep representations of EEG signals. 
Our approach is inspired by \cite{bashivan2015learning} and uses a Spatio-Spectral Feature Image (SSFI) of multi-channel EEG signals as an input to a 2D CNN model. We base our approach on the common assumption that different frequency bands in EEG signals are associated with different brain activities. We convert multi-channel EEG signals into sequences of SSFIs, consisting of a set of six images corresponding to the most widely-used six frequency bands. We train our 2D CNN model with SSFI generated from multi-channel EEG signals from the PhyAAt dataset \cite{bajaj@phyaat2020} and visualize the deep features learned at different layers as topographic maps. In addition, we carry out an Inter-Subject Dependency (ISD) analysis to explore generalized representations across subjects \cite{choi2022comparison}, \cite{lee2022inter}. 


\begin{figure*}[ht!]
    \centering
    \includegraphics[width=\linewidth]{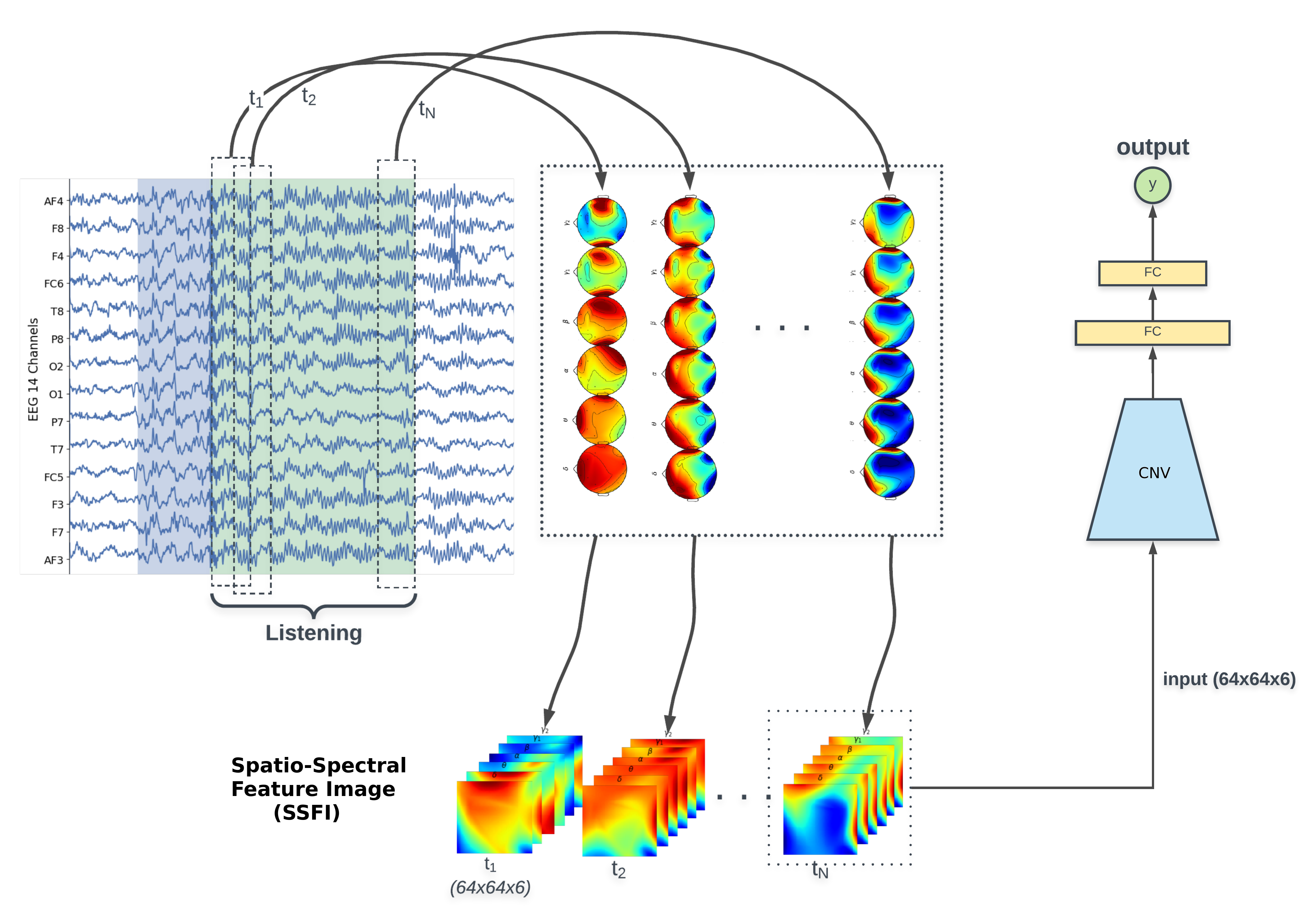}
    \caption{CNN approach with SSFI for predictive modeling}
    \label{fig:CNNAproach}
\end{figure*}

\section{Dataset}
\label{ss:dataset}
In this study we use the Physiology of Auditory Attention (PhyAAt) dataset \cite{bajaj@phyaat2020}. The PhyAAt dataset provides with a collection of 14-channel EEG signals recorded from 25 healthy subjects who underwent a total of 144 auditory attention experiments. Each experiment consisted of three tasks. First, participants were presented with an audio message reproduced under different auditory conditions (listening task); afterwards, the participants transcribed the audio message (writing task), and finally, they enjoyed a resting period before the beginning of the following experiment (rest task). The experimental auditory conditions included different levels of background noise, message lengths and message semanticity, and the transcription of each audio message was used to define an auditory attention score. An Epoc-Emotiv device \cite{EmotivEpoc} 
was used to record 14-channel EEG signals from each participant. Electrodes were arranged following the standard 10-20 EEG electrode placement, the sampling rate was set to 128 Hz and the average duration of the complete series of experiments was 40 minutes. Finally, the time periods covering a single task were labeled according to the type of task, the auditory conditions and the resulting attention score. Required ethical approval for the experiment was acquired \href{https://PhyAAt.github.io}{https://PhyAAt.github.io}. 

\section{Methods}
\label{S:Exp}
We considered the ternary classification problem of predicting whether a 1-second long 14-channel EEG segment was recorded during a listening, writing or resting task. Our proposed processing pipeline consists of two stages, as shown in Figure \ref{fig:CNNAproach}. First, 1-second long 14-channel EEG segments are extracted and transformed into SSFI arrays. Then, each SSFI array is used as an input to a CNN (trained model), which labels the segment as either listening, writing or resting. In this section, we describe the transformation of EEG segments into SSFI arrays and our CNN approach.
\begin{figure}[ht!]
    \centering
    \includegraphics[width=0.8\linewidth]{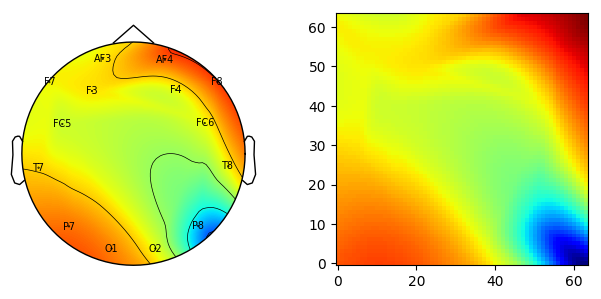}
    \caption{Spatial power distribution for a single frequency band on a scalp topographic map and its corresponding SSFI rectangular grid component. The spatial power distribution has been obtained from a 14-channel EEG signal whose recording electrodes are identified on the scalp topography. 
    }
    \label{fig:SSFI-oneband}
\end{figure}
\begin{figure*}[h!]
    \centering
    \includegraphics[width=0.9\linewidth]{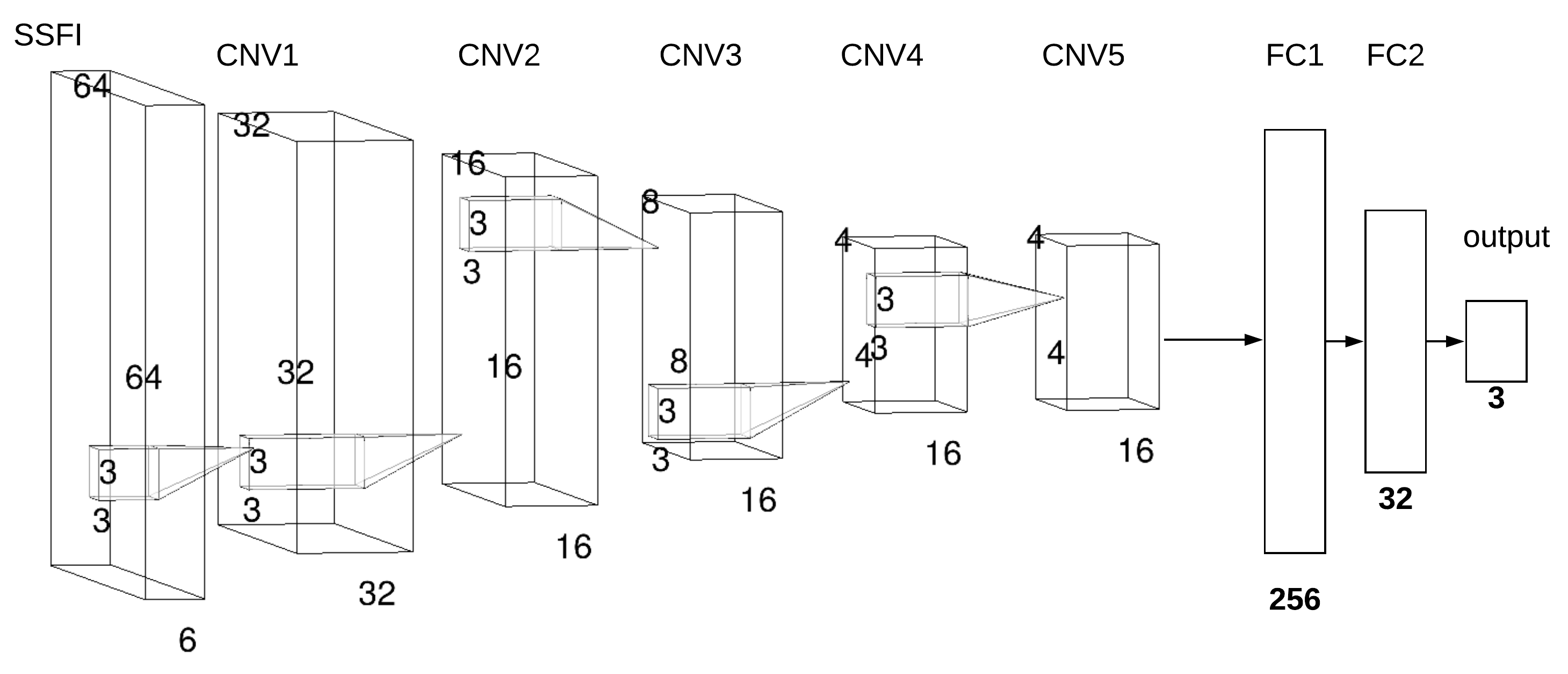}
    \caption{CNN architecture used for LWR classification}
    \label{fig:CNN}
\end{figure*}
\subsection{Spatio-Spectral Feature Image Arrays}
\label{S:SSFI}
Spatio-Spectral Feature Image (SSFI) is an image representing the spatial and spectral pattern of the data, EEG in our case. Since, in EEG studies, there are well established frequency bands associated with several mental and emotional state of the brain, rather than converting EEG data into a single SSFI, we convert them into array of different frequency bands.
We transformed raw 14-channel EEG segments into 3D SSFI arrays representing the EEG spatio-spectral distribution of power. The shape of the 3D SSFI array is $D_1\times D_2\times D_3$, where $D_1\times D_2$ is the size of a 2D grid representing the scalp topography and $D_3$ is the number of frequency bands. Preserving the scalp topography is important in multi-channel EEG studies, as spatially close scalp sites are in general affected by common brain sources.

Given a multi-channel EEG segment and a set of $D_3$ frequency bands of interest, an SSFI array is generated as follows. 
The rectangular grids are used to represent the spatial density of power within one frequency band and are constructed so that the scalp locations corresponding to the recording electrodes are associated to one of the grid locations. Based on the power spectral density of each EEG channel, the power in each frequency band is computed and assigned to the electrode's entry in the corresponding rectangular grid. The spatial density of power on the $D_1\times D_2$ grid is then obtained by interpolating and extrapolating the values computed from each electrode, by using a bicubic method adapted from the MNE library \cite{gramfort2014mne}. Figure \ref{fig:SSFI-oneband} shows the spatial density of power for a single frequency band on a scalp topography and its corresponding rectangular grid. By stacking the $D_3$ spatial densities of power, we obtain the final SSFI array of dimensions $D_1\times D_2\times D_3$.

\subsection{EEG pre-processing, feature extraction and SSFI generation}
\label{ss:ssfi_gen}
Prior to the generation of SSFI arrays, the 14-channel EEG signals were pre-processed as follows. First, each EEG channel was filtered with a $5^{th}$-order, highpass IIR filter with cut-off frequency of 1 Hz. Then artifacts were removed using ATAR Algorithm (automatic and tunable artifact removal algorithm) described in \cite{bajaj2020automatic}, with parameter $\beta=0.1$. After this pre-processing stage, 1-second long (128 samples) segments were extracted, allowing 0.75 seconds (96 samples) overlap (shift 0.25 seconds, 32 samples) between consecutive segments.

The power spectral density of each segment channel was obtained by using the Welch method with a Hamming window. Based on the estimated power spectral density, the power within the following 6 frequency bands was computed: $0.1-4$ Hz (delta, $\delta$), $4-8$ Hz (theta, $\theta$), $8-14$ Hz (alpha, $\alpha$), $14-30$ Hz (beta, $\beta$),  $30-47$ Hz (low gamma, $\gamma_1$), and  $47-64$ Hz (high gamma, $\gamma_2$). This process resulted in a feature vector $F$ of 84 ($6 \times 14$) dimensions, $F \in \mathbb{R}^{84}$, per EEG segment:
\begin{equation}
\label{eqn:XT}
F = \begin{bmatrix}
       F_{\delta}\\
       F_{\theta}\\
       F_{\alpha}\\
       F_{\beta}\\
       F_{\gamma_1}\\
       F_{\gamma_2}\\
     \end{bmatrix}  
\end{equation}

where $F_{i} \in \mathbb{R}^{14}$ is the power of the 14 channels of an EEG segment within each band $i\in \{\delta, \theta, \alpha, \beta, \gamma_1, \gamma_2\}$. A $64 \times 64$ rectangular grid was chosen for the scalp topography, resulting in a $64\times 64\times 6$ SSFI array, which was obtained as described in Section \ref{S:SSFI}. Figure \ref{fig:CNNAproach} illustrates the process of extracting a sequence of SSFI arrays from consecutive 14-channel EEG segments from the PhyAAt dataset.

\subsection{Convolutional Neural Network architecture}
\label{ss:cnn}
A CNN architecture (Figure \ref{fig:CNN}) consisting of five convolutional layers (CNV), two fully connected layers (FC) and one output layer with 3 output units was used to predict the experimental task during which a 14-channel EEG segment had been recorded.
The number of input channels in this architecture is six, which corresponds to the number of frequency bands in the input SSFI arrays. Each convolutional layer consists of a bank of filters of size $3\times 3$, followed (except for the CONV5 layer) by a $2\times 2$ max-pool layer. Each max-pool layer is followed by batch normalisation and dropout ($0.3$) layer. As shown in  Figure \ref{fig:CNN}, the number of filters for the first layer (CNV1) are 32 and 16 for the rest of the CNV layers. The activation function used for the output layer is softmax whereas for the hidden layers is a rectified linear unit (ReLu) with $l_2=0.01$ regularization parameter.

\subsection{Training and test strategy}
\label{ss:trainTesting}

Since individual differences in the brain folding structures can result in participants producing different EEG distributions \cite{IndepBrain}, we took a subject-specific approach and fitted the CNN architecture to each participant separately. In creating a training and test sets, we took into consideration the existing overlap between consecutive EEG segments, as it can potentially lead to information leak from the training to the test stage. Hence, for each subject, the EEG segments corresponding to the first 100 auditory tasks were used for training, whereas the EEG segments extracted from the remaining 44 auditory tasks were used for test. 
This serial split was used taking into consideration the temporal nature of the experiments in the PhyAAt dataset, to replicate a scenario where a machine is trained and evaluated in real-time.
We used the classification accuracy to quantify the performance of the resulting CNN models. The categorical cross-entropy was chosen as a loss function during training \cite{krizhevsky2012imagenet} and the Adam (Adaptive Moment Estimation) method was used during optimization. 

\subsection{Deep representation analysis}
\label{ss:deeprepr}
One common approach to explore a trained CNN model is to visualize the input patterns that activate trained neurons. By visualizing such pattern, it is possible to identify the primitive patterns that each layer in the CNN is looking for and explore the overall processing pipeline. 
This approach is appropriate in this study since our input image is not a conventional image of an object or human. In our case, the part of the image that activates the neurons can be visualized to investigate the learned features and representation using saliency maps \cite{simonyan2013deep} or guided backpropagation \cite{springenberg2014striving}.
A pattern is generated by feeding a random image to a trained network and maximising the activation of a selected deep neuron by optimising the input image using the gradient ascent method \cite{zeiler2014visualizing}. Since our input image is not a conventional RGB image and six channels represent the different frequency bands, we optimise and generate the pattern for each channel separately.

\subsection{Inter-Subject dependency analysis}
In order to understand the diversity of brain activity for the same task, we carried-out an ISD analysis. In ISD analysis, a trained model on one subject is tested on all the other subjects. For consistency of comparative results, a trained model is also tested on the data from the same subject, including training and testing data. This analysis allows us to understand the need for invariant features. 


\section{Results}
\label{S:results}
Based on the experiment described in Section \ref{S:Exp}, the results are analyzed in three ways. First we discuss the performance of the trained model on each individual subject. Then we use the ISD analysis to explore how each model performs on other subjects. Finally, we present the deep representations and kernel weights learned from each trained model. 
\begin{figure}[hb!]
    \centering
    \includegraphics[width=1.0\linewidth]{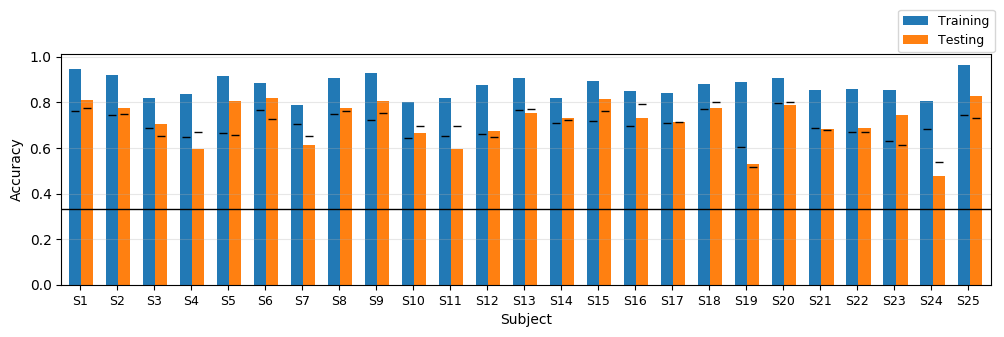}
    \caption{Results of CNN model for individual subject. A bar shows the naive model performance and a black line shows random chance level performance.}
    \label{fig:IndiSubject}
\end{figure}
\subsection{Individual subject model}
The results of CNN model, trained on an individual subject are shown in Figure \ref{fig:IndiSubject}. In addition to random chance level performance (accuracy = 1/3), a performance of a naive model (majority class) is also marked as a small bar for each subject. A naive model is considered to produce a constant output as a prediction for every input, based on the prior distribution of the class labels - majority class. Since, the time duration taken by each subject for writing and resting activity varies, the number of feature vectors (data-points), extracted from each segments also varies, producing the a different distribution of class labels. In this experiment, length of writing segments are more than listening and resting for all the subjects, thus the data-points for writing class is in majority and naive model produces label - writing for each input.

Training and test performance of the CNN models was found to be better than random chance. Compared to a naive model, CNN model performs better only for training data, however, for test data, it performs better for 16 subjects out of 25. It is interesting to observe that the same CNN model, with same fixed hyper-parameters, performs differently for each subject. By tuning a CNN model for individual subjects, the performance was improved, however to analyze the ISD associated to the task, each model was trained with a fixed parameters. 

\subsection{Inter-Subject Dependency Analysis}
\begin{figure}[h!]
     \centering
         \centering
         \includegraphics[width=0.5\textwidth]{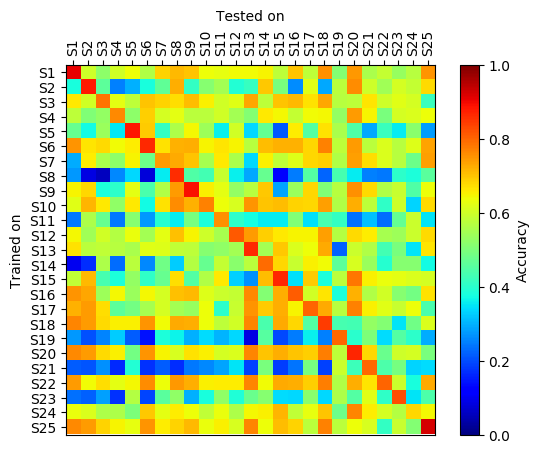}
    \caption{Inter-subject dependency analysis: Performance of a model on other subject's data. Model of trained on a subject (y-axis), tested on subject (x-axis)}
    \label{fig:ISDmat}
\end{figure}

The results of the ISD analysis are shown in Figure \ref{fig:ISD}. Figure \ref{fig:ISDmat} shows a matrix representing the performance of the CNN models trained on each subject for each subject on whom the were tested. 
The diagonal of the matrix shows the model trained and tested on the same subject, hence the accuracy is higher on the diagonal.

\begin{figure}[h!]
     \centering
     \begin{subfigure}[b]{0.5\textwidth}
         \centering
         \includegraphics[width=\textwidth]{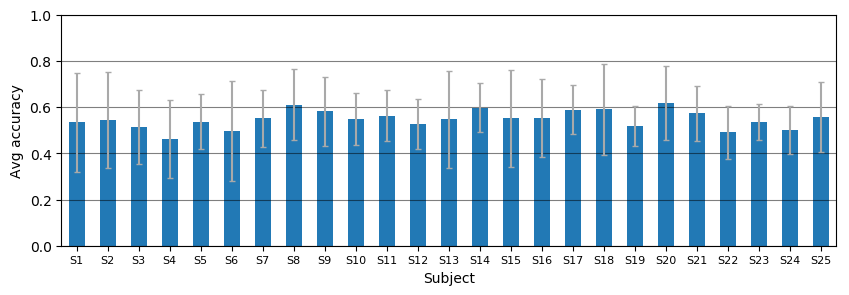}
         \caption{Subject}
         \label{fig:ISD_subj}
    \end{subfigure} 
    \begin{subfigure}[b]{0.5\textwidth}
         \centering
         \includegraphics[width=\textwidth]{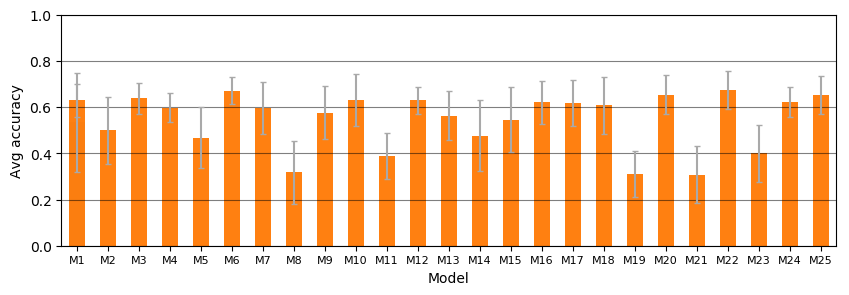}
         \caption{Model}
         \label{fig:ISD_model}
    \end{subfigure}
    \caption{The average performance of (a) subject (tested on models trained on other subjects) and (b) model (tested on other subjects).}
    \label{fig:ISD_mod_sub}
\end{figure}

The matrix shown in Figure \ref{fig:ISDmat} reveals interesting relations between subjects. First of all, it is interesting to note that the matrix is not symmetric, which could have suggested that the performance of a model trained on subject A and tested on subject B, should be similar to the performance of a model trained on subject B and tested on subject A. It can be observed that compared to other models, a CNN model trained on subject-1, perform better on other subjects, and on the other hand, a model trained on subject-19, performs poorly on other subjects.
The results from Figure \ref{fig:ISDmat} can be summarized as follows: (a) the average performance of a subject (data) on the models trained on other subject, as shown in Figure \ref{fig:ISD_subj}, and (b) the average performance of a model tested on other subjects, as shown in Figure \ref{fig:ISD_model}. 


\subsection{Deep representation and kernels}
\begin{figure}[h!]
\centering
\begin{subfigure}{0.49\textwidth}
\centering
    \includegraphics[width=1.0\linewidth]{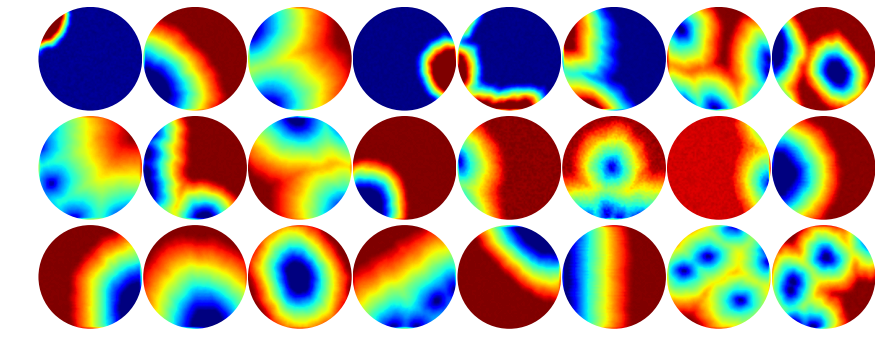}
    \caption{CNV1-layer}
    \label{fig:cnv1}
\end{subfigure}
\begin{subfigure}{0.49\textwidth}
\centering
    \includegraphics[width=1.0\linewidth]{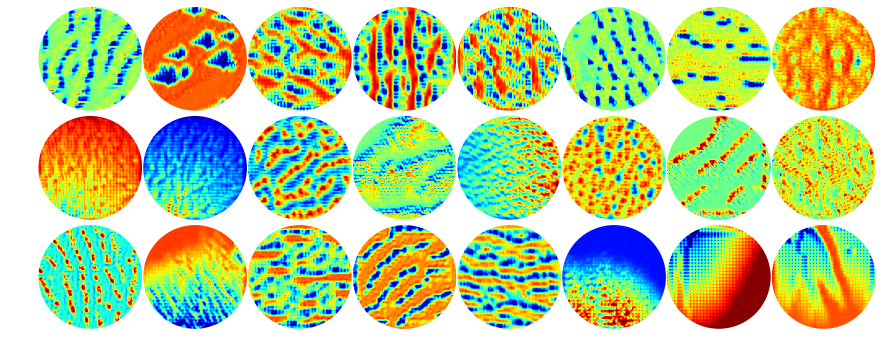}
    \caption{CNV2-layer}
    \label{fig:cnv2}
\end{subfigure}
\begin{subfigure}{0.49\textwidth}
\centering
    \includegraphics[width=1.0\linewidth]{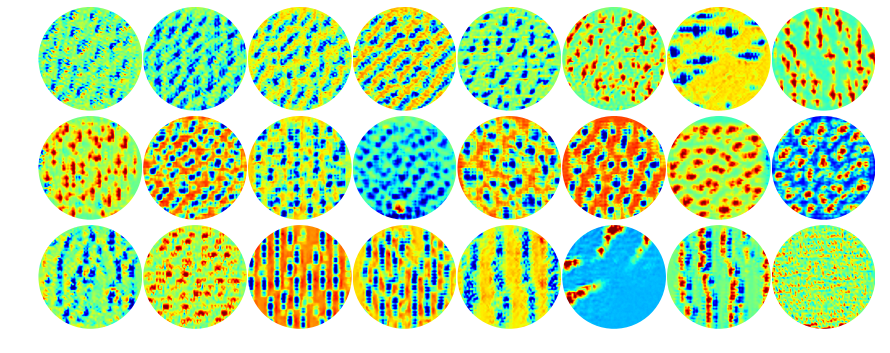}
    \caption{CNV3-layer}
    \label{fig:cnv3}
\end{subfigure}
\begin{subfigure}{0.49\textwidth}
\centering
    \includegraphics[width=1.0\linewidth]{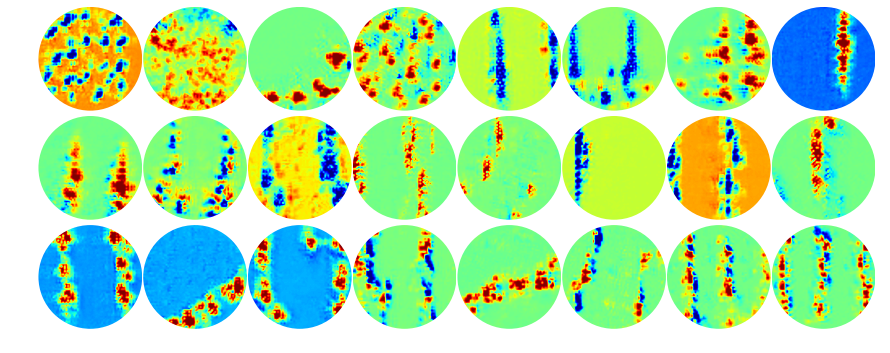}
    \caption{CNV4-layer}
    \label{fig:cnv4}
\end{subfigure}
\begin{subfigure}{0.49\textwidth}
\centering
    \includegraphics[width=1.0\linewidth]{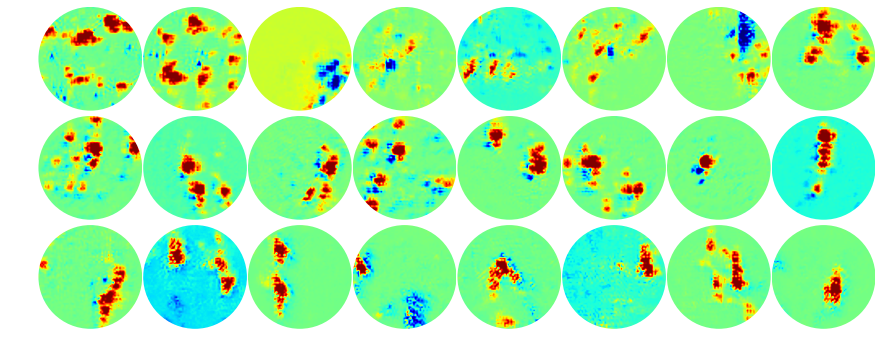}
    \caption{CNV5-layer}
    \label{fig:cnv5}
\end{subfigure}
\caption{Deep Representation}
\label{fig:DeepRepresantation}
\end{figure}

A few selected patterns generated for each CNV layer of a trained model are shown in Figure \ref{fig:DeepRepresantation}. To interpret the results, each generated pattern is displayed as a topographical image. 
It is interesting to observe that the low level features, learned from CNV1-layer (shown in Figure \ref{fig:cnv1}, are representing the the activity in different part of the Brain. For example, in left most column of Figure \ref{fig:cnv1}, pattern at the top representing the high activity in left prefrontal cortex of the Brain. Similarly, pattern at the bottom in same column shows the low level activity in right temporal and occipital lobe. Observing a bottom pattern on third column from right in same figure, represents the high activity in right brain. These pattern shows that node of CNN model gets activated when sees respective pattern. 

\begin{figure}[h!]
\centering
\begin{subfigure}{0.49\textwidth}
\centering
    \includegraphics[width=0.99\linewidth]{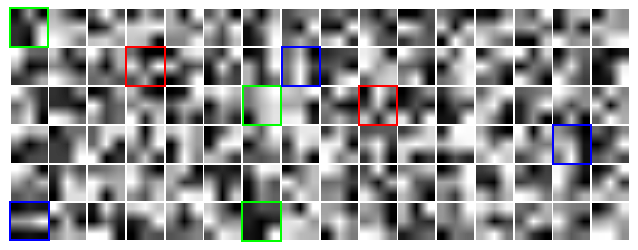}
    \caption{Kernels of CNV1-layer}
    \label{fig:cnv1k}
\end{subfigure}
\begin{subfigure}{0.49\textwidth}
\centering
    \includegraphics[width=0.99\linewidth]{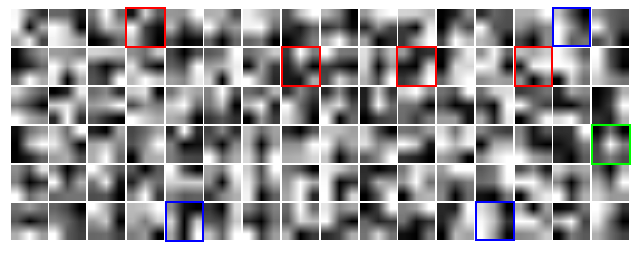}
    \caption{Kernels of CNV5-layer}
    \label{fig:cnv5k}
\end{subfigure}
\caption{Kernels learned from a CNN model at first and last layers}
\label{fig:Deepkernels}
\end{figure}



In the first layer CNV1, the features learned are smoother and represents the specific part and larger part of the Brain activity. However, as shown in Figure \ref{fig:DeepRepresantation}, the feature learned in in CNV2 and CNV3 are more focused on to smaller region. It is interesting to note that the nodes of CNV2 and CNV3 are activated with a cluster of activity in the Brain forming a grid pattern. Further, in CNV4, these grid patterns begin to shrink and focus in very specific and small patterns of activity, as shown in Figure \ref{fig:cnv4}. Interestingly, CNV5-layer, high level features, represents the activity in a very few part of Brain. For example, in Figure  \ref{fig:cnv5} bottom right pattern represent the activity in sensorimotor cortex, a middle part of Brain and top right pattern represents the activity in frontal lobe.  

It is worth mentioning that the pattern shown in Figure are not specified with respective frequency band, since we did not find any significant difference in the patterns for each band, which could be further analysed \ref{fig:Deepkernels}.



In addition to the analysis of patterns that activate node of CNN, it is worth analysing the characteristics of the kernels learned for each filter. A node in CNV layers, consists of 6 kernels corresponds to each frequency band. Figure \ref{fig:Deepkernels} shows the kernel learned for CNV1 and CNV5 layer of a trained model. As expected, the kernels learned from a model for EEG data, using SSFI do not have similar characteristics like ImageNet Neural Network \cite{krizhevsky2012imagenet}. We do not see many Gabor filters like characteristics, which are useful to find the edges in an image of a physical object. However, we can observe a few kernels as highlighted with blue boxes in Figure \ref{fig:Deepkernels}, displaying a characteristics to detect, vertical and horizontal edges. Looking into the kernels, highlighted by green boxes, they are detecting the impulsive activity in represented in SSFI, for example, a green box in CNV5, detects the activity in center of kernel, while green boxes in CNV1, detects the activity in one corner of the kernel. A peculiar characteristics shown by red boxes can be observed. 
\section{Conclusions}
\label{S:conclusion}
The approach presented in this article has the potential to exploit the spatial relationship of EEG electrodes to learn a better representation from a trained model. The analysis can be further used to derive an effective transfer learning for EEG studies. The representations learned can be further analyzed to explain the decision process of a model and to analyze the inter-subject functional correlation \cite{hasson2004intersubject}. A CNN model used for our study can also be improved by processing each channel of SSFIs separately since they are independent in terms of the brain functionality they represent.

\ifCLASSOPTIONcaptionsoff
  \newpage
\fi



%

\bibliographystyle{IEEEtran}
\bibliography{IEEEabrv,references}

\begin{IEEEbiography}[{\includegraphics[width=1in,height=1.3in,clip]{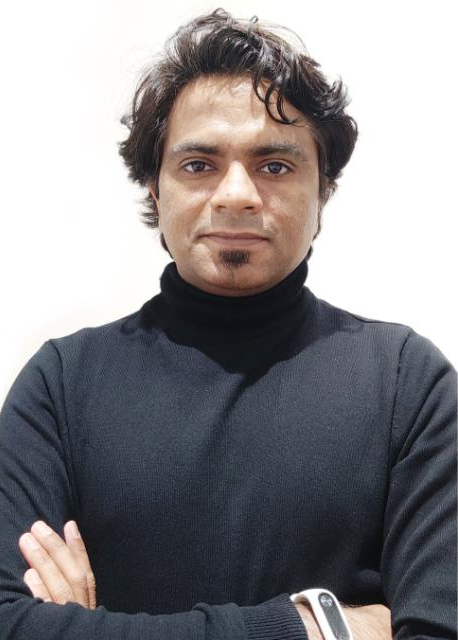}}]{Nikesh Bajaj}
completed his Ph.D. in machine learning \& signal processing from school of electronics engineering and computer science at Queen Mary University of London. UK and University of Genova, Italy, in a joint program. He is currently working as research associate at Imperial College London in National Heart \& Lung Institute (NHLI). His research work includes biomedical signal processing, machine learning, mathematical modeling, and optimization.

\end{IEEEbiography}

\begin{IEEEbiography}[{\includegraphics[width=1in,height=1.15in,clip]{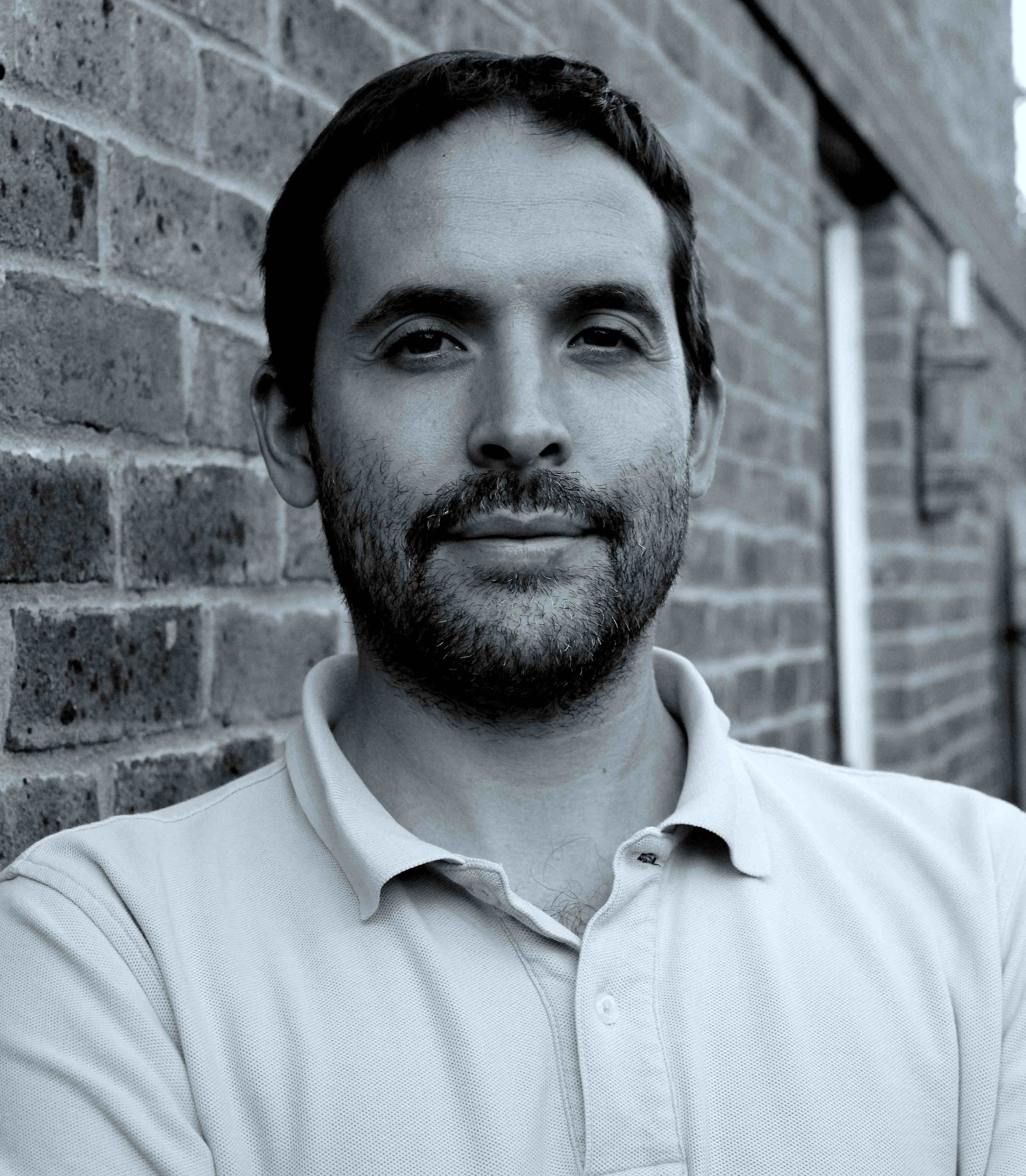}}]{Jes\'us Requena Carri\'on}
received the B.Sc.and M.Sc. degrees in telecommunications engineering
and the Ph.D. degree from the Carlos III University of Madrid, Spain, in 2003 and 2008, respectively. From 2009 to 2014, he was a Lecturer with Rey Juan Carlos University, Madrid, Spain, and in 2014, he joined the School of Electronics Engineering and Computer Science, Queen Mary University of London. He is currently a Senior Lecturer in Data Analytics at Queen Mary University of London, UK. His main research interests include statistical signal processing and computer modeling and simulation of biological systems, biomedical measurements and machine learning project management.
\end{IEEEbiography}

\begin{IEEEbiography}[{\includegraphics[width=1in,height=1in,clip]{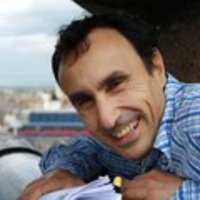}}]{Francesco Bellotti}
is currently an Assistant Professor with the Department of Electrical, Electronic, Telecommunication Engineering and Naval Architecture, University of Genova, Italy, where he teaches cyberphysical systems and entrepreneurship with the M.Sc. Program in electronic engineering. He has led WPs in several industrial research projects and has authored over 200 journals/conference papers. His main research interests
include infomobility systems, human-computer interaction, signal processing, machine learning, and technology-enhanced learning.
\end{IEEEbiography}
\vfill

\end{document}